\UseRawInputEncoding
\documentclass[preprint,12pt]{elsarticle}

\usepackage{times}
\usepackage{soul}
\usepackage{url}
\usepackage[hidelinks]{hyperref}
\usepackage[small]{caption}
\usepackage{graphicx}
\usepackage{amsmath}
\usepackage{amsthm}
\usepackage{booktabs}
\usepackage{algorithm}
\usepackage{algorithmic}
\usepackage{xcolor}
\usepackage[shortlabels]{enumitem}
\newcommand{\comment}[1]{}

\newcommand{\bw}{\mbox{\bf w}}
\newcommand{\bC}{\mbox{\bf C}}
\newcommand{\bT}{\mbox{\bf T}}
\newcommand{\bP}{\mbox{\bf P}}
\newcommand{\bV}{\mbox{\bf V}}
\newcommand{\bev}{\mbox{\bf e}}

\newcommand{\be}{\begin{equation}}
\newcommand{\ee}{\end{equation}}
\newcommand{\ZZ}{{\bf Z}}

\begin{document}

\begin{frontmatter}

\title{Fear Not, Vote Truthfully: Secure Multiparty Computation of Score Based Rules}

\author[AU,CY]{Lihi Dery \corref{CORR}}
\ead{lihid@ariel.ac.il}
\author[AU2]{Tamir Tassa}
\ead{tamirta@openu.ac.il}
\author[AU3]{Avishay Yanai}
\ead{yanaia@vmware.com}

\cortext[CORR]{Corresponding author}
\address[AU]{Department of Industrial Engineering and Management, Ariel University, Ariel, Israel}
\address[CY]{Ariel Cyber Innovation Center, Ariel, Israel}
\address[AU2]{Department of Mathematics and Computer Science, The Open University, Raanana, Israel}
\address[AU3]{VMware Research, Herzliya, Israel}

\begin{abstract}
We propose a secure voting protocol for score-based voting rules, where independent talliers perform the tallying procedure. The protocol outputs the winning candidate(s) while preserving the privacy of the voters and the secrecy of the ballots. It offers perfect secrecy, in the sense that apart from the desired output, all other information -- the ballots, intermediate values, and the final scores received by each of the candidates -- is not disclosed to any party, including the talliers. Such perfect secrecy may increase the voters' confidence and, consequently, encourage them to vote according to their true preferences. The protocol is extremely lightweight, and therefore it can be easily deployed in real life voting scenarios.
\end{abstract}


\end{frontmatter}

\newpage
\section{Introduction}\label{intro}
Ballot secrecy is an essential goal in the design of voting systems.
When voters are concerned for their privacy, they might decide to vote differently from their real preferences, or even abstain from voting altogether. Our main goal here is achieving perfect ballot secrecy.
The usual meaning of privacy in the context of secure voting is that the voters remain anonymous. Namely, even though the ballots are known (as is the case when opening the ballot box at the end of an election day), no ballot can be traced back to the voter who cast it. We go one step further and consider perfect ballot secrecy, or full privacy \citep{chaum1988}, i.e., given any coalition of voters, the protocol does not reveal any information on the ballots, beyond what can be inferred from the published results.

The mere anonymity of the ballots might not provide sufficient privacy and hence may encourage untruthful voting, as our next two examples show.
Consider a group of faculty members who need to jointly decide which applicant to accept to the faculty out of a given list of candidates.
To that end, each faculty member (voter) anonymously casts a ballot. A tallier counts the ballots and uses some voting rule to determine the elected candidate(s).
The problem with this voting strategy is that even though the tallier cannot link voters to ballots, he does see the actual ballots. Hence, besides the final outcome, say, that Alice is the elected candidate, the tallier is exposed to additional information which may be sensitive; e.g., that the candidate Bob received no votes, even though some of the voters declared upfront that they are going to vote for him. The imperfect privacy of such a voting system may cause some voters to vote untruthfully. The protocol that we present herein offers perfect privacy and, thus, may encourage
voters to vote truthfully.

As another example, consider the London Inter-Bank Offered Rate (LIBOR)\footnote{See ICE LIBOR \url{https://www.theice.com/iba/libor}} which
is the benchmark interest rate at which banks can borrow from each other. The rate is computed daily; banks that are benchmark submitters contribute to setting the LIBOR by means of voting: each bank's ``vote" is an interest rate and the LIBOR is determined by some averaging over the submitted votes.
The bank's submitted rate may signal the bank's financial viability.  Worrying about the signal which their submitted rate conveys,
some banks may submit an untruthful rate. To prevent this, the bank's individual submissions (the ballots) are kept private and are published only three months after the submission date. However, the tallier is exposed to these ballots and may be able to link some ballots to banks by financial analysis. Therefore, even anonymous ballots might not provide sufficient privacy.
Securing the ballots, as we suggest herein, means that there is less incentive to misrepresent one's ballot, and thus there is less incentive for strategic voting.

\textbf{Contributions.}
We present a secure protocol with perfect ballot secrecy to compute election results for score-based voting rules. This is achieved by employing cryptographic multiparty computation techniques.
Score-based voting rules are rules where a voter's ballot consists of scores given to each of the candidates, and the winner is the candidate with the highest aggregated score \citep{brandt2016handbook}.
This family includes rules such as {\sc Plurality}, {\sc Range}, {\sc Approval}, {\sc Veto}, and {\sc Borda}.
We follow what is known in cryptography as ``the mediated model'' \citep{Al08}, in the sense that
our protocol involves a set of talliers who perform the aggregation of ballots and compute the final voting results, but they are not allowed to access the actual ballots or other computational results such as the final scores of candidates. Our protocol is secure under the assumption that the talliers have an {\em honest majority}; that means that if a majority of the talliers are honest and do not collude, then no tallier can infer any information on the private ballots nor on
the aggregated scores of candidates.
Such perfect ballot privacy, by which the ballots and aggregated scores are not disclosed even to the talliers, may increase the voters' confidence and, consequently, encourage them to vote according to their true preferences.
As the protocol is compliant with all desired properties of secure voting systems, and is very efficient, it can be readily
implemented in real life voting scenarios.

The paper is organized as follows. In Section \ref{related} we review related work. In Section \ref{background} we provide the necessary preliminaries on score-based voting rules and on secret sharing schemes. Our protocol is presented and discussed in Section \ref{semihonest}. 
We analyze the computational and communication costs of the protocol in Section \ref{costanalysis}, discuss the protocol's compliance with essential electronic voting requirements in Section \ref{discussion}, and conclude in Section \ref{conc}.

\section{Related work}\label{related}
The issue of enhancing democratic elections is widely studied in the AI community and specifically in the computational social choice community. Some recent studies look at securing attacks during the recounting of ballots \citep{elkind_ijcai2019}, optimal attack problems in voting (e.g. by deleting voters \citep{dey_ijcai2019}), electoral bribery problem \citep{chen_ijcai2019}, 
election control through social
influence \citep{coro_ijcai2019}
and the complexity of multi-winner voting rules \citep{yang_ijcai2019}. 
A central goal in this context is the design of systems in which the election results reflect properly and truthfully the will of the individuals in the underlying society. An important vehicle towards achieving that goal is to secure the voting system so that it provides desired properties such as anonymity/privacy, fairness, robustness, uniqueness, and uncoercibility.

Previous studies on secure voting focused on different desired properties, e.g. privacy, or anonymity (a ballot cannot be connected to the voter who cast it), uniqueness (every voter can vote once), correctness (the issued winners are the ones that should be selected by the underlying voting rule from the cast ballots), and fairness (all voters must cast their ballot without seeing other votes or intermediate voting results). See e.g. 
\citet{Chang&Lee2006,gritzalis2002principles,zagorski2013remotegrity}.

Our focus is on preserving privacy and achieving perfect ballot secrecy.
One way to achieve those goals is by using methods that allow the voters to compute the outcome themselves
without relying on a tallier to aggregate and count the votes, e.g. \citep{benkaouz2015distributed}.
Another way is to use a third-party, a.k.a {\em a tallier}. In order to secure the transition of the votes which are sent from the voters to the tallier, various cryptographic techniques were utilized in prior art.

Early studies used the notions of mix-nets and anonymous channels \citep{chaum1988,ParkIK93,SakoK95}.
Blind signatures \citep{Chaum82} were used in other secure e-voting protocols, e.g. \citep{FujiokaOO92,ibrahim2003secure}.
\citet{chen2014secure} proposed a secure e-voting system based on the hardness of the discrete logarithm problem.
\citet{benaloh1987verifiable} proposed a practical scheme for conducting
secret-ballot elections in which the outcome of an election is verifiable by all
participants and even by non-participating observers; his scheme is based on secret sharing homomorphisms
\citep{Benaloh86a} that allow computations on shared data.

A large number of studies utilized homomorphic encryption, as it enables voting aggregation in the
ciphertext domain. For example, \citet{cramer1997secure}
proposed a scheme in which each voter posts a single encrypted ballot; owing to the homomorphism of the cipher,
the final tally is verifiable to any observer of the election.
\citet{damgaard2010generalization} proposed a generalization of Paillier's probabilistic
public-key system \citep{Pai} and then showed how it can be used for efficient e-voting.
While most homomorphic e-voting schemes are based on additive homomorphism, \citet{PengABDL04} proposed a scheme based
on multiplicative homomorphism. In their scheme, the tallier recovers
the product of the votes, instead of their sum, and then the
product is factorized to recover the votes.


To the best of our knowledge, only two previous studies considered the question of private
execution of the computation that the underlying voting rule dictates.
\citet{canard2018practical} considered the Majority Judgment (MJ) voting rule \citep{balinski2007theory}, which does not fall under the score-based family of rules that we consider here. They first translate the complex control flow and branching instructions that the MJ rule entails into a branchless algorithm; then they devise a privacy-preserving implementation of it using homomorphic encryption, distributed decryption schemes, distributed evaluation of Boolean gates, and distributed comparisons. \citet{NairBK15}
suggest to use secret sharing for the tallying process in Plurality voting. Their protocol provides anonymity but does not provide
perfect secrecy as it reveals the final aggregated score of each candidate. In addition, their protocol is vulnerable to cheating attacks, as it does not include means for detecting illegal votes. In our study, which covers all score-based rules, we provide perfect privacy as well as means
for preventing cheating by using a secret sharing-based secure multiparty computation (see Section \ref{Sorting}).

\section{Preliminaries}\label{background}
This section provides the required background on score based voting rules (Section \ref{votingalgs}) and secret sharing (Section \ref{ssharing}). 
\subsection{Score-based voting rules}\label{votingalgs}
We consider a setting in which there are $N$ voters, $\bV=\{V_1,\ldots,V_N\}$, that need to hold an election over $M$ candidates, $\bC = \{C_1,\ldots,C_M\}$.
The election determines a score $\bw(m)$ for each candidate $C_m$, $m \in [M]:=\{1,\ldots,M\}$ in a manner that will be discussed below.
Let $K \in [M]$ be some fixed integral parameter. Then the output of the voting algorithm is the subset of the
$K$ candidates with the highest $\bw$-scores, where ties are broken either arbitrarily or by another rule that is agreed upfront. ($K=1$ corresponds to the typical case of a single winner.)
Our protocol can be easily extended to output also the ranking of the candidates or the final scores they received. As such extensions are straightforward, we focus here on the ``lean" output consisting only of the identities of the $K$ elected candidates.

In score-based voting rules, every voter $V_n$, $n \in [N]=\{1,\ldots,N\}$, creates a ballot vector of the form $\bw_n:=(\bw_n(1),\ldots,\bw_n(M))$, where all single votes, $\bw_n(m)$, are nonnegative and uniformly bounded. Define
\be \bw = (\bw(1),\ldots,\bw(M)):= \sum_{n=1}^N \bw_n \label{wdefsum}\,.\ee
Then $\bw(m)$ is the aggregated score of the candidate $C_m$, $m \in [M]$.

We consider five types of voter inputs to be used in the above described rule template, which give rise to five well known voting rules \citep{nurmi2012comparing}:

$\bullet$ {\sc Plurality.} $\bw_n \in \{\bev_1,\ldots,\bev_M\}$ where $\bev_m$, $m \in [M]$, is an $M$-dimensional binary vector of which the $m$th entry equals 1 and all other entries are 0. Namely, $V_n$ casts a vote of 1 for exactly one candidate and a vote of 0 for all others, and the winner is the candidate who was the favorite of the maximal number of voters.

$\bullet$ {\sc Range.} $\bw_n \in \{0,1,\ldots,L\}^M$ for some publicly known $L$.
\footnote{We note that {\sc Range} is not commonly included in the family of score-based voting rules, but we include it in this family since it fits the same voting rule ``template". {\sc Range} is common in many applications, e.g. \url{www.netflix.com} and \url{www.amazon.com}, and it is often used in recommender systems \citep{masthoff2011group}.}
Here every voter gets to give a score, ranging from 0 to $L$, to each candidate.

$\bullet$ {\sc Approval.} $\bw_n \in \{0,1\}^M$. Every voter submits a binary vector in which (up to) $K$ entries are 1, while the remaining entries are 0. Such a voting rule is used when it is needed to fill $K$ equivalent positions; for example, if $K$ members in the senate of a university retired, it is needed to select $K$ new senate representatives from the faculty.

$\bullet$ {\sc Veto.} $\bw_n \in \{\hat{\bev}_1,\ldots,\hat{\bev}_M\}$
where $\hat{\bev}_m$, $m \in [M]$, is an $M$-dimensional binary vector of which the $m$th entry equals 0 and all other entries are 1.
In this method every voter states his least preferred candidate.
The winner is the candidate that got the minimal number of zero votes.

$\bullet$ {\sc Borda.} $\bw_n \in \{ (\pi(0),\ldots,\pi(M-1)): \pi \in \Pi_M \}$, where $\Pi_M$ is the set of all permutations over the set $\{0,\ldots,M-1\}$.
Here, the input of each voter is his own ordering of the candidates, i.e., $\bw_n(m)$ indicates the position of $C_m$ in $V_n$'s order, where a position of $0$ (resp. $M-1$) is reserved to $V_n$'s least (resp. most) favorite candidate.

\subsection{Secret sharing}\label{ssharing}
Secret sharing methods \citep{Shamir79} enable distributing a secret among a group of participants. Each participant
is given a random share of the secret so that: (a) the secret can be reconstructed only by combining the shares given to specific {\em authorized} subsets of participants, and (b) combinations of shares belonging to unauthorized subsets of participants reveal zero information on the underlying secret.

The notion of secret sharing was introduced, independently, by \citet{Shamir79} and \citet{Blakley}, for the case of threshold secret sharing. Assuming that there are $D$ participants, $\bP=\{P_1,\ldots,P_D\}$, then the access structure in Shamir's and in Blakley's schemes
consists of all subsets of $\bP$ of size at least $D'$, for some $D' \leq D$. Such secret sharing schemes are called $D'$-out-of-$D$.

Shamir's secret sharing scheme works as follows. Assume that $p$ is a sufficiently large prime so that the domain of all possible secrets may be embedded in the finite field $\ZZ_p$. Denote the secret to be shared by $x$. The Shamir's scheme has the following two procedures: {\sf Share} and {\sf Reconstruct}:

\smallskip
$\bullet$ ${\sf Share}_{D',D}(x)$.
    The procedure samples a uniformly random polynomial $g(\cdot)$ over $\ZZ_p$, of degree $D'-1$, where the free coefficient is $x$. That is, $g(t)=x+\alpha_1 t + \alpha_2 t^2 + \ldots + \alpha_{D'-1} t^{D'-1}$, where $\alpha_j$, $1 \leq j \leq D'-1$, are selected uniformly at random from $\ZZ_p$.\footnote{Note that the actual degree of the polynomial could be less than $D'-1$, if $\alpha_{D'-1}=0$, but for simplicity we relate to such polynomials as having degree $D'-1$.} The procedure outputs $D$ values, $x_1,\ldots,x_D$, where $x_d=g(d)$ is the share given to $P_d$, $d \in [D]=\{1,\ldots,D\}$.
    
\smallskip
$\bullet$ ${\sf Reconstruct}_{D'}(x_1,\ldots,x_D)$. The procedure is given any selection of $D'$ shares out of $\{ x_1,\ldots,x_D \}$, say $\{ x_{j_1},\ldots,x_{j_{D'}} \}$ where $1 \leq j_1 < \cdots< j_{D'} \leq D$, and it then interpolates a polynomial $g(\cdot)$ of degree at most $D'-1$ such that $g(j_i)=x_{j_i}$ for all $i \in [D']$. The procedure then outputs $x=g(0)$. 
It is easy to see that any subset of $D'-1$ (or less) shares reveals nothing about the secret $x$, whereas any subset of $D'$ (or more) shares fully determines the polynomial $g$, and in particular, the secret $x=g(0)$.

We conclude this crash course on secret sharing with the observation that the secret sharing procedure is linear in the following sense.
Let $x$ and $y$ be two secrets from $\ZZ_p$ and $a,b \in \ZZ_p$ be two publicly known values. Assume that $(x_1,\ldots,x_D)$
and $(y_1,\ldots,y_D)$
are shares in a Shamir's $D'$-out-of-$D$ secret sharing scheme
in $x$ and $y$, respectively. Then, as can be readily verified,
$(ax_1+by_1,\ldots,ax_D+by_D)$ are shares in a Shamir's $D'$-out-of-$D$ secret sharing scheme
in $ax+by$. Indeed, if $g_x$ and $g_y$ are the share-generating polynomials of degree $D'-1$ that were used to create the shares in $x$ and $y$, respectively, then the set of shares
$(ax_1+by_1,\ldots,ax_D+by_D)$ correspond to the share-generating polynomial $f:=ag_x+bg_y$ which is a polynomial of degree $D'-1$ for which $f(0)=ax+by$. 

Our protocol involves a {\em distributed} third party, $\bT=\{T_1,\ldots,T_D\}$, called the tallier ($\bT$) or talliers ($T_d$, $d \in [D]$).
In the protocol, we use secret sharing for creating shares of the private ballots of the voters and distributing them among the $D$ talliers.
As those ballots are vectors (see Section \ref{votingalgs}), 
the secret sharing is
carried out for each entry independently, so that each of the talliers receives a share vector in each ballot.

\comment{
Our protocol uses both types of threshold secret sharing schemes, over some finite field $\ZZ_p$ which is determined upfront.
Each voter $V_n$ shares his ballot vector $\bw_n$ with the talliers using the AON scheme. At the end of the voting period, each of the talliers can add the shares which it
had received from all voters in order to get AON shares in the aggregated score vector $\bw$ (see Eq. (\ref{wdefsum})).
Before moving on to the determination of the winning candidates, the talliers must translate those shares in $\bw$ into shares in a Shamir's $D'$-out-of-$D$ secret sharing scheme, with $D'=\lfloor (D-1)/2 \rfloor$. We will discuss later on how such a translation is carried out and
why it is needed.
}


\section{The method: A secure protocol for score based rules}\label{semihonest}
In this section we present our protocol. 
As indicated earlier, our protocol is mediated, in the sense that it assumes a set of talliers,
$\bT = \{ T_1, \ldots,T_D\}$, who assist in the computations, but are not allowed to learn any information on the private votes of the voters.
The number of talliers, $D$, can be any integer $D > 1$. Higher values of $D$ will imply higher computational and communication costs, but they will also imply greater security against coalitions of corrupted talliers.

A privacy-preserving implementation of
score-based rules is described in Protocol \ref{algvote}.
Before delving into that protocol, we make the following observation.
For each of the five score-based rules, there is a known upper bound $B$ on the entries of $\bw$. $B=N$ in {\sc Plurality}, {\sc Approval}, and {\sc Veto}
rules, $B=NM$ in {\sc Borda}, and $B=NL$ in {\sc Range}.
Let $p$ be a fixed prime greater than $B$. Then all computations in Protocol \ref{algvote} are carried out in the field $\ZZ_p$.

First, each voter $V_n$, $n \in [N]$, constructs his own ballot vector (Step 1), $\bw_n \in (\ZZ_p)^M$.
We assume that all voters know the index $m \in [M]$ of each candidate. For example, that index can be determined by the lexicographical ordering
of the candidates according to their names.

In Step 2, $V_n$ creates $D$ random share vectors of his ballot vector, $\bw_n$, using Shamir's secret sharing scheme with the threshold
$D'=\lfloor (D+1)/2 \rfloor$. The reason for selecting this specific threshold will be clarified later on.
The sharing is done on each entry of $\bw_n$ independently. Namely, for each $m \in [M]$, $V_n$ generates a random polynomial $g_{n,m}$ of degree $D'-1$ over $\ZZ_p$, where $g_{n,m}(0)=\bw_n(m)$. Then, in Step 3, $V_n$ sends the share vector $\bw_{n,d} = (g_{n,1}(d),\ldots,
g_{n,M}(d))$ to $T_d$, for all $d \in [D]$.

After receiving the ballot shares from all voters, $T_d$ computes the sum of all received share vectors $\hat{\bw}_d = \sum_{n=1}^N \bw_{n,d}$ mod $p$ (Step 4).
Each such vector $\hat{\bw}_d$ on its own carries no information regarding the votes (since it is the sum of uniformly random and independent vectors). But
in view of Eq. (\ref{wdefsum})
and the linearity of the secret sharing operation (see Section \ref{ssharing}), the set
$\{ \hat{\bw}_d(m): d \in [D]\}$ is a set of $D$ shares in $\bw(m)$ by a Shamir $D'$-out-of-$D$ secret sharing scheme, for all $m \in [M]$, where $\bw$ is the aggregated vector of scores. In other words, for every $m \in [M]$, there exists a polynomial $g_m$ of degree $D'-1$ over $\ZZ_p$, where $g_m(0)=\bw(m)$,
and $g_m(d)=\hat{\bw}_d(m)$ for all $d \in [D]$.

The heart of the protocol is in Step 5: here, the talliers engage in a secure multiparty computation (MPC) in order to find the indices of the $K$ candidates with the highest aggregated scores. This is a non-trivial task since no one holds the vector $\bw$.
In Section \ref{Sorting} we explain the notion of MPC and describe the MPC protocol that we use for the above described task. Once those indices are found, the voters proceed to find the identity of the $K$ candidates behind those indices (Step 6).

\floatname{algorithm}{Protocol}
\begin{algorithm}[h!!]
\caption{\label{algvote} A basic protocol for secure score-based voting}
\textbf{Input}: $\bw_n$, $n \in [N]$; $K \in [M]$.\\
\textbf{Output}: The $K$ candidates from $\bC$ with highest aggregated scores in $\bw = \sum_{n=1}^N \bw_n$.\\
\begin{algorithmic}[1]
\STATE Each voter $V_n$, $n \in [N]$, constructs his ballot vector $\bw_n$ according to the selected indexing and voting rule.
\STATE Each voter $V_n$, $n \in [N]$, generates a random polynomial $g_{n,m}$ of degree $D'-1$ over $\ZZ_p$, where $g_{n,m}(0)=\bw_n(m)$, $\forall m \in [M]$. Then, he creates the share vector $\bw_{n,d} = (g_{n,1}(d),\ldots,
g_{n,M}(d))$.
\STATE $V_n$, $\forall n \in [N]$, sends $\bw_{n,d}$ to $T_d$, $\forall d \in [D]$.
\STATE $T_d$, $\forall d \in [D]$, computes $\hat{\bw}_d = \sum_{n=1}^N \bw_{n,d}\mod p$.
\STATE $T_1,\ldots,T_D$ find the indices of the $K$ candidates in $\bC$ with highest $\bw$-scores and output them.
\STATE The voters find the identities of the top $K$ candidates.
\end{algorithmic}
\end{algorithm}

\subsection{Sorting shared vectors}\label{Sorting}
The main challenge in Step 5 of Protocol \ref{algvote} is to
find the indices of the $K$ largest entries in $\bw$.
Towards that end, the talliers can implement any sorting algorithm on $\bw$ until all $K$ largest entries are found.
However, the talliers must not reconstruct $\bw$'s entries, nor even learn any piece of information about them.
They must perform oblivious comparisons using only the shares that they hold in $\bw$'s entries.

Assume that the two entries that need to be compared are $u=\bw(m)$ and $v=\bw(m')$ for some $m,m' \in [M]$. Each tallier $T_d$, $d \in [D]$, holds
random shares $u_d,v_d \in \ZZ_p$ in $u$ and $v$, respectively, in a Shamir's $D'$-out-of-$D$ secret sharing scheme.
The talliers wish to find whether $u<v$, but without revealing any information beyond that on $u$ and $v$.

To privately verify such questions, we use a secure multiparty computation (MPC) protocol \citep{yao}.
An MPC protocol allows $T_1,\ldots,T_D$ to compute any function $f$ over private inputs $x_1,\ldots,x_D$ that they hold, so that at the end of the protocol everyone learns $f(x_1,\ldots,x_D)$ {\em but nothing else}.\footnote{Of course, some information may be inferred from the desired output $f(x_1,\ldots,x_D)$, but this is inevitable and allowed. For example, if the desired output is a median of $x_1,\ldots,x_D$, then at the completion of the protocol, every tallier whose input is smaller than the median can infer that there are at least $\frac{D}{2}$ talliers that hold greater values than his own.}
A common approach towards designing efficient MPC protocols is to represent the function $f$ by an arithmetic circuit $C$ such that for every set of inputs, $x_1,\ldots,x_D$, the output of the circuit, $C(x_1,\ldots,x_D)$, equals $f(x_1,\ldots,x_D)$.

The circuit is composed of input and output gates such that each input gate is fed with a single secret value by one of the parties, and an output gate determines a single value that is revealed to all parties.
Additionally, between the input and output gates there are multiple layers of arithmetic gates that connect them.
An arithmetic gate can be either addition or multiplication.
Each gate is given exactly two inputs, and it produces one output such that the output of a gate at layer $\ell$ can be given as input to multiple gates in layer $\ell+1$. (All input gates constitute the first layer of the circuit.) Only the value that the output gate issues is revealed to the parties; all intermediate values that pass from one gate to another remain secret from everyone. Specifically, a secure protocol allows the parties to maintain the invariant that the actual value output from each gate is secret-shared, as described in Section \ref{ssharing}. When reaching the output gate, each party broadcasts the corresponding share that it holds, so that everyone can reconstruct the output. 


The computational and communication costs of computing such circuits depend mainly on the number of multiplication gates and on the number of layers in the circuit, as we proceed to explain.
To compute a multiplication gate of two secrets, $s_1$ and $s_2$, the parties have to interact; i.e., each party needs to send some information to the other parties. However, to compute a multiplication gate of one secret $s$ and a public value $c$, the parties do not need to interact (such a gate requires only local computation). The same holds for an addition gate. Therefore, for the efficiency of secure computation, circuit designers are mostly concerned with the number of multiplication gates in the circuit and with the {\em depth} of the circuit, i.e., the number of interactions that have to be performed sequentially (since they depend on each other and cannot be performed in parallel).

Specifically, in this work we use a design of an arithmetic circuit by
\citet{NO07}, which performs an MPC comparison of two secret values $u,v \in \ZZ_p$.
It is assumed
that each of the interacting parties, $T_d$, $d \in [D]$, holds shares $u_d$ and $v_d$ in $u$ and $v$, respectively, in a Shamir's $D'$-out-of-$D$ secret sharing scheme, where
$D' \leq \lfloor (D+1)/2 \rfloor $.
The circuit
outputs the bit that indicates whether $u<v$.
The circuit has a {\em constant depth} (15 to be concrete). This is advantageous as the depth does not depend neither on the number of parties nor on the field size $p$, so changing those parameters does not have a significant effect on performance (see Table \ref{tbl:mpc-results} in Section \ref{costanalysis}).

\subsection{The protocol's security}\label{overallsecurity}
Here we discuss the security of the whole protocol.
An important goal of secure voting is to provide anonymity; namely, it should be impossible to connect
a ballot to the voter who cast it.
Protocol \ref{algvote} achieves that goal since each cast ballot is distributed into random shares and then each share is sent to a different
tallier.
Each such share carries zero information on the underlying ballot.
Even subsets of $D'-1 = \lfloor (D-1)/2 \rfloor$ shares reveal no information on the secret ballot. 
Our working assumption is that the set of talliers
has an {\em honest majority}; that assumption means
that even if some of the talliers are dishonest and try to collude in order to extract sensitive information on the ballots, the number of colluding (dishonest) talliers is smaller than the number of the honest talliers. Since each ballot is shared by a $D'$-out-of-$D$ secret sharing scheme, with $D'=\lfloor (D+1)/2 \rfloor$, then the number of talliers that have to collude in order to recover the private ballots is at least $D' \geq D/2$, and that scenario is impossible under the honest majority assumption. Hence, under that assumption the talliers cannot recover the ballots, nor can they infer even partial information on them. 

As for the MPC computation that the talliers carry out in Step 5 of Protocol \ref{algvote}, its security is proven in
\citep{NO07}. 
By utilizing that protocol, the talliers may find the indices of the $K$ winning candidates without learning any information beyond the order that the aggregated scores induce on the candidates\footnote{The circuit that we use in order to verify inequalities may be modified in order to hide intermediate comparison results and output only the $K$ indices of candidates with highest scores. Such a version, which we do not describe herein, will output only the $K$ winning candidates without disclosing their order.}. Also the security of that computation
(in similarity to the security of the secret sharing of the individual ballots) is guaranteed under the assumption of an honest majority.

Hence, to summarize, the voters' privacy is perfectly preserved by our protocol, unless at least $D'=\lfloor (D+1)/2 \rfloor$ talliers betray the trust vested in them.
For example, with $D=3$ talliers, at least two talliers would need to collude in order to recover the ballots; similarly, if $D=5$ at least three talliers would need to collude for that purpose. If such a collusion does not occur, as implied in settings with an honest majority, the talliers will be able to compute the final election results without learning {\em anything} beyond those computed results.

A collusion scenario that threatens the security of our protocol is highly improbable, and its probability decreases as $D$ increases. Ideally, the talliers would be parties that enjoy high level of trust within the organization or state in which the elections take place, and whose business is based on such trust. Betraying that trust may incur devastating consequences for the talliers. 
Hence, even if $D$ is set to a low value such as $D=5$, in which case at least 3 talliers need to collude in order to recover the personal ballots,
the probability of such a breakdown of trust in any conceivable application scenario (with a proper selection of the talliers) would be negligible.

Another possible attack scenario is as follows: a voter $V_j$ can eavesdrop on the communication link between another voter $V_n$ and each of the talliers, and intercept the messages that $V_n$ sends to the talliers (in Protocol \ref{algvote}'s Step 3) in order to recover $\bw_n$ from them; additionally, $V_j$ may replace $V_n$'s original messages that carry shares of $\bw_n$ with other messages (say, ones that carry shares of
$\bw_j$, or any other desired fake ballot).
Such an attack can be easily thwarted by requiring each party (a voter or a tallier) to
have a certified public key,
encrypt each message that he sends out using the receiver's public key and then sign it using his own private key;
also, when receiving messages, each party must first verify them using the public key of the sender and then send a suitable
message of confirmation
to the sender.
Namely, each message that a voter $V_n$ sends to a tallier $T_d$ in Step 3 of Protocol \ref{algvote} should be signed with $V_n$'s private key and then encrypted by $T_d$'s public key; and $T_d$ must acknowledge its receipt and verification.

In view of the above discussion, the tradeoff in setting the number of talliers $D$ is clear: higher values of $D$ provide higher security since more talliers would need to be corrupted in order to breach the system's security. However, increasing $D$ has its costs: more independent and reputable talliers are needed, and the communication and computational costs of our protocol increase, albeit modestly (see Section \ref{costanalysis}).

A fundamental assumption in all secure voting systems that rely on fully trusted talliers (that is, talliers who receive the actual ballots from the voters) is that the talliers do not misuse the ballot information and that they keep it secret.
In contrast, our protocol significantly reduces the trust vested in the talliers as it denies the talliers access to the actual ballots. Even in scenarios where some (a minority) of the talliers betray that trust, privacy is ensured. Such a reduction of trust in the talliers is essential in order to increase the confidence of the voters in the voting system so that they would be further motivated to exercise their right to vote and moreover, vote according to their true preferences, without fearing that their private vote would be disclosed to anyone.

\subsection{Validating the legality of the cast ballots}\label{validating}
Protocol \ref{algvote} is designed for honest voters, namely, voters who cast legal votes.
However, voters may attempt cheating by submitting illegal ballots in order to help their candidate of choice.
For example, assume that $V_n$'s favorite candidate is $C_m$ and the voting rule is {\sc Plurality}.
Then a honest $V_n$ would cast the ballot $\bw_n=\bev_m$ (see Section \ref{votingalgs}).
A dishonest $V_n$, on the other hand, could cast the ballot $\bw_n=N \bev_m$. Such an illegal ballot would boost $C_m$'s chances of winning,
or, if $V_n$ is the only dishonest voter, it would even ensure $C_m$'s win. Similar options of cheating exist also with the other voting rules. 
Since the talliers do not see the actual ballots, if a voter can pull such a cheat, it might remain undetected.

In real-world voting scenarios, where voters typically cast their ballots on certified computers in voting centers, the chances of hacking such computers and tampering with the software that they run are small. However, for full-proof security, we proceed to describe an MPC solution
that enables the talliers to validate the legality of each ballot, even though those ballots remain hidden from them.
In case a ballot is found to be illegal, the talliers may recover it (by adding up all shares) and use the recovered ballot as a proof
of the voter's dishonesty.

Let us start by examining the {\sc Plurality} rule.
A ballot $\bw_n$ is legal in this case iff $\bw_n(m)\leq 1$, i.e. if $\bw_n(m)\in\{0,1\}$) for all $m \in [M]$, and $\sum_{m \in [M]} \bw_n(m)=1$ mod $p$ (assuming $p>M$). 
Each of the above $m$ inequalities 
can be verified by an MPC sub-protocol that computes an arithmetic circuit that outputs the product $\bw_n(m)\cdot (\bw_n(m)-1)$; a suitable MPC protocol that we may adopt in our context is described in \citet{ChidaGHIKLN18}. The talliers accept the vote $\bw_n(m)$ as legal (namely, being
either 0 or 1) iff the result is 0. 
Finally, verifying that there exists exactly one entry in the vector $\bw_n$ with the value 1 is done by computing the sum $\sum_{m \in [M]} \bw_n(m)$ and verifying that it equals 1. The fact that $N<p$ ensures that there will not be a wrap around.

To validate ballots $\bw_n$ in the case of {\sc Veto} or {\sc Approval}, we also need to check that each entry, $\bw_n(m)$, $m \in [M]$, is either 0 or 1, as described above for {\sc Plurality}. If all entries were validated, the talliers need to proceed and check an aggregated condition on the ballot's entries.
The aggregated condition in {\sc Veto} can be checked by computing the sum $\sum_{m \in [M]} \bw_n(m)$ and verifying that it equals $M-1$. The aggregated condition in {\sc Approval} requires that
the sum $S_n:=\sum_{m \in [M]} \bw_n(m)$ is at most $K$.
That condition can be checked by a circuit that outputs
$S_n\cdot (S_n-1) \cdots
(S_n-K)$. The talliers will accept $\bw_n$ as legal iff
the latter product equals 0. Indeed, if that product equals 0 then the talliers can deduce that $V_n$ had voted for at most $K$ candidates, without knowing the exact number of candidates for whom $V_n$ had voted. Any result other than 0 will testify that $V_n$ had cheated; in such cases the talliers can reject his vote and consider further consequences.

In the case of {\sc Range}, a ballot $\bw_n$ is legal iff $\bw_n(m)\leq L$ for all $m \in [M]$.
Each of the these $m$ inequalities can be verified by a circuit that outputs $\bw_n(m)\cdot (\bw_n(m)-1) \cdots
(\bw_n(m)-L)$.
The talliers will accept the ballot $\bw_n$ iff the result of that circuit will be 0 for each $m \in [M]$.

The {\sc Borda} rule is slightly different. A ballot is legal under this rule if it consists of some permutation of the $M$ values in $\{0,1,\ldots,M-1\}$. Hence, a ballot is legal iff it satisfies the following
two conditions:
\be \bw_n(m) \leq M-1 ~~~ \forall m \in [M] \,;~~\mbox{and} \label{bordaX1}\ee
\be \bw_n(m) \neq \bw_n(m') ~~~\forall m > m' \in [M] \,. \label{bordaX2}\ee
The entry-wise conditions in Eq. (\ref{bordaX1}) can be verified as described earlier.
The global condition in Eq. (\ref{bordaX2}) can be verified by verifying that each of the ${M \choose 2}$ differences
$\bw_n(m) - \bw_n(m')$, $m > m' \in [M]$, is nonzero. However, for privacy reasons, the talliers must not recover those differences, since they would reveal the entire ballot. To prevent such leakage of information and still allow the verification, the talliers can generate ${M \choose 2}$ random secret elements in
 the field $\ZZ_p$, denoted by $\rho_{m.m'}$, $m > m' \in [M]$, and then compute
$\zeta_{m,m'} :=\rho_{m,m'} \cdot ( \bw_n(m) - \bw_n(m'))$.
If the underlying field $\ZZ_p$ is large, then with high probability (of $1/p$) the selected $\rho_{m.m'}$ is nonzero. In such cases, $\zeta_{m,m'} \neq 0$ iff $\bw_n(m) - \bw_n(m') \neq 0$, and $\zeta_{m,m'}$ reveals no information at all on
$\bw_n(m) - \bw_n(m')$ (since $\rho_{m,m'}$ can be any nonzero element in the underlying field).

Generating shares of a random secret multiplier is a simple task. In fact, such a protocol is executed by the talliers anyway as part of the secure computation protocol of  \citet{ChidaGHIKLN18}. Specifically, for securely computing a multiplication gate in that protocol, the talliers generate two random secret values in the field. Therefore, in our computational cost analysis (Section \ref{costanalysis}), we upper bound the cost of generating shares in a random secret by the cost of evaluating a multiplication gate (where in fact the latter cost is strictly higher than the former).

Lastly, we consider the case of false negatives. In the field $\ZZ_p$ there is a probability of $1/p$ that the resulting random element would be zero. In such a case, $\zeta_{m,m'}$ would be zero even though $\bw_n(m) - \bw_n(m') \neq 0$. Hence, the talliers would get a false alarm regarding a ballot of some voter, as if it contains two equal entries, when in fact all of the ballot's entries are distinct, as required. But the probability of such a false alarm is $1/p$. In Section \ref{costanalysis} we present runtimes for the field $\ZZ_p$ with $p=2^{61}-1$; in such large fields the probability $1/p$ is negligible. Furthermore, even if for some difference
$\bw_n(m) - \bw_n(m')$ the talliers get an indication that it equals zero, they can repeat the test with another random and independent multiplier $\rho_{m.m'}$. If even that additional test yields $\zeta_{m,m'}=0$, the talliers can withold that ballot until its validity is verified (say, by performing additional independent tests until the probability of an error reduces to below some given threshold, or by revealing the value of the difference $\bw_n(m) - \bw_n(m')$).

\comment{
The global condition in Eq. (\ref{bordaX2}) can be verified by computing a circuit that outputs
\be P_n:=\prod_{m > m' \in [M]} \left( \bw_n(m) - \bw_n(m') \right)\,.  \label{prodborda}\ee
The ballot is legal iff $P_n$ is nonzero. It can be easily verified that if the ballot is legal, namely, it consists of a permutation of the values in $\{0,1,\ldots,M-1\}$, then $P_n$ will equal $\sigma \cdot C_M$ where $C_M:=\prod_{k=1}^{M-1} k!$ and $\sigma$ is the sign of the permutation in $\bw_n$.
Hence, by computing the product in Eq. (\ref{prodborda}), the talliers will not learn anything on the ballot except for the sign of the underlying permutation. 
To hide even that information, the talliers can compute $(P_n)^2$, instead of $P_n$, and verify that it is nonzero. 
By doing so, the validation process of legal ballots would not reveal any information on the ballot, beyond its validity. 
}

\comment{
We note that the number of multiplications in the circuit that computes $P_n$ is quadratic in $M$, while the value of $C_M$ is exponential in $M$. However, the Borda rule is relevant only for cases where $M$ is small, say $M \leq 10$, since for higher values of $M$ it is not realistic to expect voters to have a full ranking on all candidates.}

\section{Evaluation: Computational and communication costs}\label{costanalysis}
We analyze herein the computational and communication costs for the voters (Section \ref{costv}), and for the talliers, where the latter discussion is separated to two parts --- the
cost for computing the final election results (Section \ref{costt1}) and the cost for validating the legality of the cast ballots (Section \ref{costt2}).

\subsection{Costs for the voters}\label{costv}
The costs for each voter are negligible, as a voter needs only to generate $M(D-1) \log_2 p$ random bits, perform $M(D-1)$ additions in $\ZZ_p$,
and then send $D$ messages of $M \log_2 p$ bits each (Protocol \ref{algvote}, Steps 2-3).

\subsection{The talliers: the cost of computing the final election results}\label{costt1}
The cost in Step 4 of Protocol \ref{algvote} is negligible ($(N-1)M$ additions in $\ZZ_p$), but the determination of the winners (Step 5) is more costly as it invokes a protocol for secure comparison. 
The number of multiplication gates in the comparison circuit is $279\cdot\log p +5$ in a circuit of depth 15. 
A secure evaluation of a multiplication gate incurs a communication of $12\log p$ bits per tallier (according to \citet[Table 2]{ChidaGHIKLN18}). Hence,
the overall communication per tallier 
for a single comparison
is roughly $3348\cdot\log^2 p$ bits (or $< 1.5$ megabytes when $p=2^{61}-1$).

In order to evaluate the runtime of performing such a comparison, we ran it on Amazon AWS {\sf m5.4xlarge} machines at N. Virginia over a network with bandwidth 9.6Gbps.
We performed our evaluation with $D\in\{3,5,7,9\}$ talliers.
As for the bound $B$ on aggregated scores, which affects the runtime, we examined two cases: $B < p_1$ and $B <p_2$, where $p_1:=2^{13}-1$ and $p_2:=2^{31}-1$ are two Mersenne primes. Namely, in cases where there are few voters, or ballots' entries are small, and as a result $B<p_1$, we used the bound $p_1$; otherwise, we used the bound $p_2$, which seems to suffice for all conceivable application scenarios.
Using Mersenne primes is advantageous in the context of secure computation, since multiplication of two field elements can be done without performing an expensive division. The results are presented in the first two rows of Table \ref{tbl:mpc-results}. 

Note that those runtimes are for a single comparison. In order to determine the identity of the $K$ winners, it is necessary to perform up to $KM$ comparisons. We can see that even in large election scenarios that require choosing an underlying field of size $p_2$, with $D=9$ talliers (for enhanced security), with large numbers of candidates, $M$, and selected winners, $K$, the election results can be determined within only few seconds. 

\begin{table}[ht]
	\small
	\centering
	\setlength\tabcolsep{4.2pt} 
	\begin{tabular}{l | l | l | l | l}
		\hline
		 		 	& $D=3$ 	 & $D=5$  		& $D=7$  	& $D=9$\\
		\hline
		$B<p_1=2^{13}-1$	& 2.83	 & 4.3 	& 	6.6	& 12.81	\\
		\hline
		$B<p_2=2^{31}-1$	&  9.07	 & 9.54	& 	9.64	& 15.0	\\
		\hline
		\hline
		Validating $5\cdot 10^4$ {\sc Plurality} ballot entries & 41.3 & 42.2 & 52.9 & 65.55 \\
		\hline
		\hline
		Validating $5\cdot 10^4$ {\sc Range} ballot entries with $L=20$ & 826 & 844 & 1058 & 1311 \\
		\hline
		Validating $5\cdot 10^4$ {\sc Range} ballot entries with $L=100$ & 
		4210& 4945& 5770& 7050 \\
		\hline
		\hline
		Validating $5\cdot 10^4$ {\sc Borda} {ballots} with $M=5$ &  1652	 & 1688 	& 	2116	& 2622 \\
		\hline
		Validating $5\cdot 10^4$ {\sc Borda} {ballots} with $M=10$ & 
		7434	 & 7596 	& 	9522	& 11799 \\
\hline
	\end{tabular}
	\caption{Rows 1-2 show runtimes (milliseconds) for a secure comparison protocol with a varying number of talliers, $D$, and two field sizes, $p_1$ and $p_2$. Rows 3-7 show runtimes for a batch validation of $5\cdot 10^4$ ballot entries or full ballots for various voting rules.}
	\label{tbl:mpc-results}
\end{table}

\subsection{The talliers: the cost of validating ballots}\label{costt2}
For ballot validation, we consider three rules --- {\sc Plurality}, {\sc Range}, and {\sc Borda}. (The validation of ballots in the two remaining rules, {\sc Veto} and {\sc Approval}, is similar to that in {\sc Plurality}.)

We extrapolated the runtime for executing the validation procedure for the various rules
from the experimental results reported by \citet{ChidaGHIKLN18}. 
They experimented on a similar network setting, but used a larger field with $p=2^{61}-1$. They experimented with a circuit
that consists of one million multiplication gates that are evenly spread over 
$\{20,100,1000\}$ layers; hence, in each layer there are $\{5\cdot 10^4, 10^4, 10^3\}$ multiplication gates, respectively. 
The reported runtimes as a function of $D$, the number of parties (talliers in our case), are shown in Table \ref{tbl:Chida-results}.

\begin{table}[ht]
	\small
	\centering
	\setlength\tabcolsep{4.2pt} 
	\begin{tabular}{c | c | c | c | c | c}
		\hline
		 	\#layers & \#multiplication gates per layer 	& $D=3$ 	 & $D=5$  		& $D=7$  	& $D=9$\\
		\hline
		20 &  50000 & 826	 & 844 	& 	1058	& 1311	\\
		\hline
		100 & 10000 	& 842	 & 989 	& 	1154	& 1410\\
		\hline
		1000 & 1000 	& 1340	 & 1704 	& 	1851	& 2243\\
		\hline
	\end{tabular}
	\caption{Runtimes (milliseconds) for computing $10^6$ multiplication gates, spread evenly over 20, 100, and 1000 layers, as a function of the number $D$ of talliers.
	The first two columns show the number of layers and the number of multiplication gates per layer in each setting.}
	\label{tbl:Chida-results}
\end{table}

\subsubsection{The {\sc Plurality} rule}
As described in Section \ref{validating}, a {\sc Plurality} ballot validation depends on a circuit that checks whether $\bw_n(m) \leq 1$.
As explained there, such a circuit requires only one multiplication gate and has depth 1. In addition, to verify that there is only one ballot entry with the value 1, we need only to perform summation,
which requires no multiplication gates.
Overall, a validation circuit of $q$ ballots requires exactly $qM$ multiplication gates in {\em one layer}.
As explained earlier, the runtimes in row 1 of Table \ref{tbl:Chida-results} are for a circuit with $10^6$ multiplication gates
that were spread evenly over 20 layers. Hence, the runtime of executing
$10^6/20 = 5 \cdot 10^4$ multiplication gates in a single layer is obtained by dividing those runtimes by 20, as shown in row 3 of Table \ref{tbl:mpc-results}.
Those are the runtimes for Validating $5\cdot 10^4$ {\sc Plurality} ballot entries (from several different voters).
In fact, those runtimes constitute an upper bound on the actual runtimes for our application, since they were obtained with
a field size of $p=2^{61}-1$ that is larger than $p_1$ or $p_2$ that suffice for our needs.
Those numbers indicate that validation is an extremely lightweight task in the case of the {\sc Plurality} rule (as well as {\sc Veto} and {\sc Approval}). Validating 50 million ballot entries can be done in roughly one minute.

\comment{
Overall, a validation circuit for the entire {\sc Plurality} ballot requires exactly $M$ multiplication gates.
We extrapolated the runtime for executing the validation circuit from the experimental results reported by \citet{ChidaGHIKLN18}. In fact, their results constitute an upper bound on the runtime required by our application. Specifically, they experimented over a similar network setting (over AWS EC2 machines), but used a larger field with $p=2^{61}-1$. For a circuit of depth 20 and $10^6$ multiplication gates they report runtimes of $\{826, 844, 1058, 1031\}$ milliseconds when the number $D$ of parties (talliers in our case) is $\{3,5,7,9\}$, respectively. Since a validation of a single entry in a ballot has only one multiplication gate, we can safely divide their runtimes by 20 and obtain the runtime of Validating $10^6/20=5\cdot 10^4$ {\sc Plurality} ballot entries (possibly of several different voters). The results appear in the third row of Table \ref{tbl:mpc-results}.
}

\subsubsection{The {\sc Range} rule}
Here we evaluate the cost of validating a ballot in the {\sc Range} rule, as a function of the maximum score, $L$, that a voter can give to a candidate. The validation of an entry in a {\sc Range} ballot requires a circuit with $L$ multiplications in a row (when implemented na\"{i}vely, without optimizations). Thus, the experiment in \citet{ChidaGHIKLN18} with a circuit that consists of a million multiplication gates spread over 20 layers (the runtimes of which are reported in row 1 of Table \ref{tbl:Chida-results}) captures exactly $5\cdot 10^4$ {\sc Range} ballots entries, in the case $L=20$. We report those numbers in row 4 of
Table \ref{tbl:mpc-results}.
When $L$ equals $100$, we rely on the runtimes that are reported in \citet{ChidaGHIKLN18} for a circuit of 
a million multiplication gates spread over 100 layers, as appear in row 2 of
Table \ref{tbl:Chida-results}.
In terms of validating {\sc Range} ballot entries, those are the runtimes for validating 10000 {\sc Range} ballot entries with $L=100$. We multiply those runtimes by 5, in order to get the same batch size as in the case of $L=20$, and report the resulting runtimes in row 5 row of Table \ref{tbl:mpc-results}.

We infer that even though validating {\sc Range} ballot entries, especially for large values of $L$, is more costly than validating {\sc Plurality} ballot entries, the validation task remains a very practical one. For example, validating 50 million {\sc Range} ballot entries, even for $L=100$ and $D=9$, can be done in under two hours. (Recall that the election process usually spans a long period, say one day. Hence, the validation process can be spread along the entire election period, by validating each time the batch of ballots that were cast since the last validation.)

\subsubsection{The {\sc Borda} rule}
Finally, we turn to evaluate the cost of validating a ballot in the {\sc Borda} rule, as a function of the number of candidates, $M$.
As described in Section \ref{validating}, validating a {\sc Borda} ballot consists of two stages. In the first stage we check that each ballot entry is in the range $[0,M-1]$. This can be done by a circuit of depth $M-1$ with an overall number of $M(M-1)$ multiplication gates. In the second stage we check that the ballot is a permutation of
 $\{0,\ldots,M-1\}$.
This verification consists of generating ${M \choose 2}$ secret random sharings and a secure computation of ${M \choose 2}$ multiplication gates (see Section \ref{validating}) in a circuit with a single layer.
The computation and communication required for generating shares of a secret random value are strictly smaller than those required for a multiplication gate, since in the protococl of \citet{ChidaGHIKLN18}, evaluating a multiplication gate involves the generation of two secret random sharings. Thus, 
the cost of the second stage in the validation of a {\sc Borda} ballot
can be bounded by the cost of performing $2\cdot {M \choose 2}=M(M-1)$ multiplications, in two consecutive layers. In total, the overall cost of both stages in validating a single {\sc Borda} ballot can be bounded by performing $2M(M-1)$ multiplications over $M-1$ layers.\footnote{Note that by taking a spread of the needed workload over more layers than what is actually needed, one gets higher runtimes. We perform that relaxation for the sake of simplicity, since we are only interested in upper bounds on the validation cost that will demonstrate its lightweight nature.}

When $M=5$, the overall cost of validating a single ballot can be bounded, as stated above, by performing $2M(M-1)=40$ multiplications spread over 4 layers. Hence, to validate a batch of $2.5 \cdot 10^4$ {\sc Borda} ballots in this case we need to perform one million multiplications spread over 4 layers. Those runtimes, for the various values of $D$, can be bounded by the runtimes in row 1 of Table \ref{tbl:Chida-results}, since the latter runtimes are for the same number of multiplications, only spread over more layers (20 instead of 4). Finally, we conclude that the overall runtime for validating a batch of $5 \cdot 10^4$ {\sc Borda} ballots is given by doubling the runtimes in row 1 of Table \ref{tbl:Chida-results}, as shown in row 6 of 
Table \ref{tbl:mpc-results}.

Bounding the runtimes for $M=10$ is done in a similar fashion.
The overall cost of validating a single ballot is smaller than that of performing $2M(M-1)=180$ multiplications spread over 9 layers. Hence, to validate a batch of $10^6/180$ {\sc Borda} ballots in this case we need to perform one million multiplications spread over 9 layers. Those runtimes, for the various values of $D$, can be bounded by the runtimes in row 1 of Table \ref{tbl:Chida-results}. Finally, we conclude that the overall runtime for validating a batch of $5 \cdot 10^4$ {\sc Borda} ballots is given by multiplying the runtimes in row 1 of Table \ref{tbl:Chida-results} by 9, as shown in row 7 of 
Table \ref{tbl:mpc-results}.

In cases where $M \geq 22$, we can use the runtimes reported in rows 2 or 3 of Table \ref{tbl:Chida-results}, since in such cases the number of layers (in the first stage), which is $M-1$, would be greater than 20. However, as the {\sc Borda} rule is impractical for such values of $M$, because it is hard to expect voters to have a full ranking over such a large number of candidates, we leave it for the interested reader to compute the resulting runtime bounds for such values of $M$. But the numbers for $M=5$ and $M=10$ indicate that the validation of {\sc Borda} ballots is a very practical task, despite the higher complexity of that task in comparison to that in {\sc Plurality} or {\sc Range}. For example, the validation of one million ballots when there are $M=10$ candidates and $D=9$ talliers would take less than 4 minutes.

\section{Discussion}\label{discussion}
Electronic voting schemes should, ideally, comply with some essential requirements. Below we list those requirements, as defined in \citet{Chang&Lee2006}. We discuss, in a theoretical manner, the compliance of our proposed protocol with each of the requirements.
(Such a discussion of compliance with essential requirements is common in papers dealing
with secure electronic voting schemes, see e.g. \citet{chung2009approach,li2008electronic,liaw2004secure,wu2014electronic}.)

{\bf Anonymity/Privacy.} {\em The votes should remain anonymous throughout the entire process.} Our protocol achieves anonymity, as explained in Section \ref{overallsecurity}. In particular, our protocol outputs only the desired election results, without revealing any information on the ballots.

{\bf Fairness.} {\em A fair mechanism does not reveal intermediate results.} Our protocol is fair since all intermediate results (such as aggregated scores of candidates) are kept secret-shared; the protocol is designed to output just the identities of the winner(s) and nothing beyond that.

{\bf Convenience.} {\em No special equipment is required and the voters do not need to learn any specialized technique.} Our protocol relies on basic and general-purpose cryptographic functions that can be found in many free libraries.
In fact, as our protocol includes a mechanism for validating the cast ballots (Section \ref{validating}), voters may vote from anywhere (and not necessarily from voting centers).
They only need to download a simple software package that implements our protocol vis-a-vis the talliers.

{\bf Robustness.} {\em No malicious intruder should be able to interrupt the procedure.} This requirement is addressed by the standard security mechanisms that should be implemented on top of our protocol, as discussed in Section \ref{overallsecurity}: signing ballots before sending, verifying received ballots, and confirming the receipt and verification of each ballot.

{\bf Mobility.} {\em The mechanism can be implemented to run on the World Wide Web.} Our protocol allows mobility since it requires no special equipment and it relies on common cryptographic toolkit.

{\bf Uniqueness.} {\em Each voter is allowed to vote only once.} If a voter attempts to vote twice (Step 3 in Protocol \ref{algvote}), only his first vote will be processed, while the second one will be ignored.
Also, one voter cannot shoot down a message from another voter (in order to prevent the latter voter's ballot to reach the talliers) since the voter expects to receive from each tallier a confirmation of receiving his signed and authenticated message.
In case such a confirmation is not received, the voter can resend his message.

{\bf Completeness.} {\em Only eligible voters are allowed to vote.} This requirement is achieved by the usage of certificated public keys
and signatures.

{\bf Uncoercibility.} {\em A voter must not be able to prove to a third party how he had voted, in order to prevent bribery.}
The only way to prove to a third party $W$ how a voter
$V_n$ had voted is if all $D$ talliers send to $W$ the shares of $V_n$'s ballot $\bw_n$.
Hence, under our assumption that the tallier has an honest majority, our protocol offers uncoercibility.

{\bf Correctness.} {\em Each ballot must be counted correctly.} Correctness follows directly from the talliers' semi-honesty (namely, that they follow the prescribed computations correctly). To ensure the talliers' semi-honesty in practical deployments, the talliers' software must be verified and authenticated, and it would be best to run it on a dedicated and tamper-proof machine (namely, a computer that is physically protected from any hacking attempts).

{\bf Efficiency.} {\em The computational load of the whole process is required to be such that the result is obtained within a reasonable amount of time.}
Our protocol complies with this requirement, as is evident from our cost analysis that shows that it is
extremely lightweight
(see Section \ref{costanalysis}).

{\bf The right to abstain.} {\em A voter that wishes to abstain should be able to do so without revealing that fact to any other party \citep{gritzalis2002principles}}.
In all considered rules, each ballot is a vector
and the final result is determined by the sum of the private vectors. Hence, a voter $V_n$ who wishes to abstain can use $\bw_n={\bf 0}$. Such an action is equivalent to not participating.
Even the talliers cannot tell that a message received from some voter contains an all-zero vote, because of the secret sharing.
Hence, our protocol allows confidential abstinence.

\bigskip
In view of the above, our proposed scheme is suitable for  electronic elections of any kind and scale, from
small-scale elections similar to the examples given in Section \ref{intro}, to national elections in populations of any size, provided that the underlying voting rule is score-based. In national elections, it is possible to implement our electronic voting system in a manner that resembles existing voting systems. 
Citizens will arrive at the voting centers and identify using their standard identification card. They will then obtain access to a computer into which they will privately enter their selection. The software running the protocol will translate their selection into a ballot vector, as described in Section \ref{votingalgs}; subsequently, shares of that ballot will be distributed to the $D$ talliers' servers, who will then proceed to process them as described in Section \ref{semihonest}. As opposed to existing tallying systems, in our system, the {\it final and true} results will be determined with utmost certainty and security within seconds after the voting period ends. 

\section{Conclusion}\label{conc}

We begin this section with a birds-eye summary of our study. Then, we identify the limitations of our method and provide an outlook on future research directions.

\subsection{Summary}
We considered a setting in which a group of voters wishes to elect $K$ candidates out of a given list of candidates. We considered score-based rules and showed how to securely compute the winners using a secure multiparty protocol.
Our protocol offers perfect ballot secrecy:
apart from the desired output (the identity of the elected $K$ candidates),
all other information, such as the actual ballots or the aggregated scores that
each of the candidates received, is kept secret from all parties -- voters and talliers alike. 
(As indicated in Section \ref{overallsecurity}, the protocol herein allows the talliers to deduce the ranking among the candidates, but a small modification of the underlying circuit may hide even that information.)

Such level of privacy may be essential in some scenarios. For example,
when a prize committee needs to select $K$ recipients out of $M$ candidates, it is desirable to determine only
the identity of the $K$ prize recipients
without revealing their internal ranking or their aggregated scores, as such pieces of information might expose undesired information about the ballots.

Such perfect secrecy may increase the confidence of the voters in the voting system, so that they would be encouraged to participate in the elections and vote truthfully without fearing that their private vote would be disclosed to anyone else.
Furthermore, our technique can be used during iterative voting \citep{airiau2009iterated,lev2016convergence,meir2010convergence,dery2019lie}. If, during iterative voting,
the candidates' scores are kept secret from all parties, then voters would not obtain from the voting process information that could have been used for strategic voting. (Of course, the perfect secrecy of our protocol cannot stop strategic voting altogether, since voters may still base strategic voting on information from polls, rumors and other communication channels.)

The protocol complies with conventional security desiderata,
as discussed in Section \ref{discussion}. 
An analysis of its computational and commutation costs shows that it is practical as it is extremely lightweight. 
The protocol is based solely on existing cryptographic arsenal.
This is a prominent advantage of our protocol; indeed, protocols that can be implemented on top of existing libraries are advantageous over protocols that require the development, scrutiny and
assimilation of new cryptographic components and, therefore, might be unattractive to practitioners.

\subsection{Limitations and future work}
Our method has several limitations, which suggest corresponding future research endeavours to resolve them.

\begin{enumerate}[(a)]
    \item \textbf{Convincing the public.} As with each new technology, our model faces the difficulty of ``selling" it to the public and the legislators, and convincing them of its safety and other desired advantages. As a first step in this direction, a wide-range user study should be carried out.  
    \item \textbf{Extension to order-based rules.} Herein we focused on score-based rules. 
    Another important family of rules consists of rules where the voter submits an ordered list of preferences. 
    Such rules are called {\em order-based} or {\em pairwise-comparison} voting rules (see \citet{brandt2005decentralized}). Two prominent rules in this family are -- {\sc Copeland} \citep{copeland1951reasonable}
    and {\sc Maximin} (a.k.a {\sc Kramer-Simpson}) \citep{kramer1977dynamical,simpson1969defining}. We intend to devise a protocol for securely computing the winners in voting systems that are based on such rules.
    \item \textbf{Extension to multi-winner elections.} Our current protocol can select the top $K$ candidates, for any $K<M$; namely, it can output the $K$ candidates that got the highest aggregated scores. However, there are multi-winner election protocols that are designed specifically for selecting the $K$ candidates that would satisfy the voters the most \citep{faliszewski2017multiwinner, elkind2017multiwinner}, in the sense that they also comply with additional social conditions (e.g., that the selected winners include a minimal number of representatives of specific gender, race, region etc.). This problem has unique features and it therefore requires its own secure protocol. Examples for voting rules that are designed for such a purpose are {\sc Chamberlin-Courant} \citep{chamberlin1983representative} and {\sc Monroe} \citep{monroe1995fully}.
    \item \textbf{A hierarchical tallier model.} 
    We assumed a ``flat'' tallier model, where all talliers are operating on all ballots. However, in large voting systems, a hierarchical tallier may be more suitable. For example, in the US, it may be more suitable to use a hierarchy by county (first level), state (second level), and national (third and highest level). A modification of our protocol for such settings is in order.
\comment{
    \item \textbf{Robustness.}
    In our protocol, all ballots are shared between all $D$ talliers using a simple $D$-out-of-$D$ secret sharing scheme (see Section \ref{ssharing}). As a result, our protocol is very secure, since the ballots are perfectly secure, unless all $D$ talliers collude and recover the ballots by combining all of their shares. However, another implication of using such a secret sharing scheme is that even if a single tallier, say $T_D$, becomes dysfunctional (e.g., due to a physical damage to its servers, or a cyberattack) then all ballot information is lost and the voting procedure should be performed again with $T_1,\ldots,T_{D-1},T'_D$, where $T'_D$ is another new tallier, or $T'_D$ is $T_D$ after it had recovered. In order to add robustness to our system, a Shamir $D'$-out-of-$D$ secret sharing scheme (Section \ref{ssharing}) should be used instead. In that case, even if any subset of $D-D'$ talliers become dysfunctional, then the remaining $D'$ talliers can still complete the computation of the final voting results. In order to do that, our MPC protocol should be modified in order to be compliant with such secret sharing.
}
    \item \textbf{Restraining malicious talliers.} 
    Our protocol assumes that the talliers are semi-honest,
    i.e., they follow the prescribed computations correctly. As discussed in Section \ref{discussion}, the semi-honesty of the talliers can be ensured in practice by securing the software and hardware of the talliers. However, one can provide even a stronger protection shield by cryptographic means. Namely, it is possible to design an MPC protocol that would be immune even to malicious talliers that may attempt to infer secret ballot information or intermediate voting results, or may even try to affect the final voting results. MPC protocols that are designed to be immune to malicious parties are usually significantly costlier and more complex than the corresponding MPC protocols for semi-honest parties. However, since in common voting systems, the final results are published usually hours and even days after voting had ended, the implementation of costlier MPC protocols that are designed to resist malicious talliers is not expected to be problematic, and it will enhance even further the security of the system and the trust of voters in the preservation of their privacy.  
\end{enumerate}

\bigskip
To conclude, the secure voting protocol that we presented here is of relevance to any setting in which voting should be implemented in a manner that ensures perfect privacy.
The protocol promotes truthful voting and the final results are obtained within seconds.
Even though the presented protocol is relevant and adequate as is to a wide range of practical voting scenarios, it opens up several interesting research directions that could widen its application scope and further strengthen its security.


\section*{Funding}
Lihi Dery was supported by the Ariel Cyber Innovation Center in
conjunction with the Israel National Cyber directorate of the Prime
Minister’s Office. The authors hereby declare that this support did not
influence the work reported in this paper and that the funding source
had no involvement in the project.


\end{document}